\documentclass[prb,preprint]{revtex4-1} 
\usepackage{amsmath}  
\usepackage{amsfonts} 
\usepackage{graphicx} 
\usepackage[dvipsnames]{xcolor}
\usepackage{ulem} 

\begin{document}

\title{Space pirates: a pursuit curve problem involving retarded time}


\author{Thales Azevedo}
\email{thales@if.ufrj.br}
\affiliation{Instituto de F\'\i sica, Universidade Federal do Rio de Janeiro, 
Rio de Janeiro, RJ, Brazil}

\author{Anderson Pelluso}
\email{pelluso14@gmail.com}
\affiliation{Instituto de F\'\i sica, Universidade Federal do Rio de Janeiro, 
Rio de Janeiro, RJ, Brazil}


\date{\today}

\begin{abstract}
We revisit the classical pursuit curve problem solved by Pierre Bouguer in the 18th century, taking into account that information propagates at a finite speed. To a certain extent, this could be seen as a relativistic correction to that problem,  though one does not need  Einstein's theory of relativity in order to derive or understand its solution. 

The discussion of this generalized problem of pursuit constitutes an excellent opportunity to introduce the concept of retarded time without the complications inherent to the study of electromagnetic radiation (where it is usually seen for the first time), which endows the problem with a clear pedagogical motivation.

We find the differential equation which describes the problem, solve it numerically, compare the solution to Bouguer's for different values of the parameters, and deduce a necessary and sufficient condition for the pursuer to catch the pursued, complementing previous work by Hoenselaers.
\end{abstract}

\maketitle 

\section{Introduction}

 In 1732, French mathematician,  geophysicist and 
hydrographer Pierre Bouguer posed and solved a problem which is nowadays regarded as the beginning of modern mathematical pursuit analysis. \cite{Pierre} The problem consisted of finding the functional form of the curve described by a pirate ship in pursuit of a merchant vessel, subject to the conditions that both move at constant speeds and, at any given time, the velocity vector of the pirate ship points in the direction of the merchant vessel position at that time. This situation is illustrated in figure~\ref{fig.original} (section~\ref{sec:original}).

Bouguer's original problem has been analyzed through a number of different approaches, as well as generalized to various other cases (see, for example, refs. \cite{Boole,Bernhart,Bernhart2,Behroozi,Colman,Mungan,Silagadze}), and even a three-dimensional version has been considered. \cite{Barton3d} An inspiring introduction to Bouguer's and other pursuit problems is given in a nice book by Paul Nahin. \cite{Nahin}

Though irrelevant in most cases of interest, there is a clear physical inconsistency in all the above-mentioned problems: they tacitly assume that information on the position of the merchant vessel reaches the pirate ship instantly, i.e. that whatever signal used by the pirates to infer the merchants' position (such as light or sound) travels at infinite speed. Nonetheless, it is a well-known consequence of Einstein's theory of relativity that nothing can travel faster than the speed of light in empty space. \cite{Schutz}

To the best of our knowledge, that issue has only been addressed in a relatively recent paper by Hoenselaers, \cite{Hoenselaers} in which the author considers a relativistic correction to the classical (pure) pursuit problem. However, in our opinion, the relativistic aspect of the problem is overemphasized in that paper, potentially scaring away readers who are not familiar with the theory. Indeed, although ultimately based on the postulates of special relativity, the assumption of a finite speed of propagation is all one needs in order to analyze the problem in a more physically accurate manner.
In fact, as alluded to in the previous paragraph, one could imagine a situation in which visibility is too low and the pirates are guided only by the sounds produced by the merchants. In that case, for speeds comparable to the speed of sound, the corrections analyzed here would become very important, whereas relativistic effects would play no role at all. 

The purpose of the present paper is therefore to complement the analysis done in ref. \cite{Hoenselaers}, solving the problem in a slightly different, more intuitive way. In particular, we emphasize the concept of retarded time, which is crucial to the solution of the problem. We believe that introducing that concept in the context of this (essentially classical mechanical) problem, without the complications inherent to the study of electromagnetic radiation (where it is usually seen for the first time), will be beneficial to students, which endows the problem with a clear pedagogical motivation.
Moreover, we include graphs comparing the trajectories of the pirate ship in both versions of the problem, as well as the derivation of a necessary and sufficient condition for the pirates to reach the merchants, which were absent in Hoenselaers's paper. We also discuss in some detail to what extent the problem analyzed in this paper can be regarded as a relativistic correction to the original one.

\section{Review of the original problem\label{sec:original}}

In this section, we review the original problem posed by Bouguer, following closely the solution given in ref. \cite{Nahin}.
Consider a merchant vessel traveling at constant speed $V_m$ along some given (known) trajectory. Not so far from it, a pirate ship traveling at constant speed $V_p$ is in pursuit of that vessel, following a curved path such that its velocity vector (tangent to the curve) always points directly towards the merchant vessel.

\begin{figure}[h!]
    \centering
    \includegraphics[width=6.5cm]{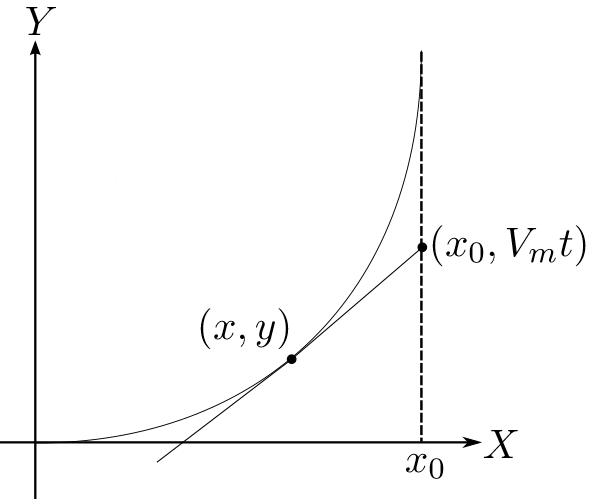}
    \caption{Sketch of the classical problem of pursuit. The figure shows the  positions of the pirate ship and the merchant vessel at a generic instant $t>0$, given respectively by $(x,y)$ and $(x_0,V_m t)$. Note that the tangent line to the pursuit curve at the point $(x,y)$ passes through the point $(x_0, V_m t)$.
}
    \label{fig.original}
\end{figure}

In the original problem, the merchants move along a straight line. Therefore, one can choose a reference frame such that at time $t \in \mathbb{R}$ the merchant vessel is located at the point $(x_0,V_m t)$, $x_0 >0$. This implies that when $t=0$ the position of the merchant vessel is given by $(x_0,0)$, and we choose the origin of our coordinate system at the pirate ship position at that time. Figure~\ref{fig.original}  illustrates this situation for some $t > 0$, when the pirate ship is located at an arbitrary point $(x,y)$. The problem consists of finding the equation defining the curved path followed by the pirate ship in the form $y=f(x)$.

 As can be seen from figure \ref{fig.original}, the slope of the tangent line to the pursuit curve at the point $(x,y)$ is given by
\begin{equation}
    \frac{dy}{dx} = \frac{V_{m}t - y}{x_{0} - x}\,.
    \label{eq.class.slope}
\end{equation}

\noindent Besides, since its speed is constant, we know that the distance the pirate ship has sailed along the pursuit curve from the origin to the point $(x,y)$ is equal to $V_p t$. Now, from calculus we know that distance is precisely the arc-length,  given by the expression below:
\begin{equation}
    V_{p}t = \int _{0}^{x} \sqrt{1+\left(\frac{dy}{d\xi} \right)^{2}}d\xi\,,
    \label{eq.class.arc}
\end{equation}
where $\xi$ is a dummy variable of integration. Combining (\ref{eq.class.slope}) and (\ref{eq.class.arc}) so as to get rid of the parameter $t$, we arrive at
\begin{equation}
\frac{1}{V_{p}}\int_{0}^{x}\sqrt{1+\left( p(\xi)\right)^{2}}d\xi = \frac{y}{V_m} - \left(\frac{x-x_0}{V_m}\right) p(x)\,,
\end{equation}
where we have defined $p(x):=dy/dx$. Differentiating the above equation  with respect to $x$, we obtain a first-order differential equation for $p(x)$:
\begin{equation}
    \frac{1}{V_{p}}\sqrt{1+p(x)^{2}} =  - \left(\frac{x-x_0}{V_m}\right)\frac{dp}{dx} \,.
    \label{eq.class.p}
\end{equation}

Equation (\ref{eq.class.p}) can easily be  integrated, given the initial condition $p(0)=0$ (cf. figure \ref{fig.original}). One finds
\begin{equation}
    p(x) =  \frac{1}{2}\left[ \left( 1 - \frac{x}{x_0}\right)^{-\alpha} - \left( 1 - \frac{x}{x_0}\right)^{\alpha}\right],
\end{equation}
where we have defined $\alpha:= V_m/V_p$. Finally, since $p(x)=dy/dx$, we can once again integrate this equation  to find 
\begin{equation}
   y = \frac{\alpha \,x_0}{1 - \alpha^2}+ \frac{x_0}{2}\left[ \frac{(1 - \frac{x}{x_0})^{1+\alpha}}{1+\alpha} - \frac{(1 - \frac{x}{x_0})^{1-\alpha}}{1-\alpha}\right],
   \label{eq.class.sol}
\end{equation}
which is the solution to Bouguer's problem. It is important to note that, in deriving this solution, we have assumed $\alpha<1$, i.e. $V_m < V_p$. If $\alpha\geq1$, then the expression for $y$ is such that it diverges as $x$ approaches $x_0$, implying that the pirates never reach the merchants.

\section{Finite speed of propagation and retarded time}\label{s.retarded}

In Bouguer's problem, it is tacitly assumed that the pirates perceive any change in the position of the merchant vessel instantaneously. In other words, information is assumed to propagate at infinite speed. As mentioned in the introduction, that would be in conflict with Einstein's theory of relativity.
In this section, we revisit the problem solved in section~\ref{sec:original}  taking into account that information must propagate at a finite speed. This will naturally lead us to the concept of retarded time. Apart from that, we follow more or less the same steps as in the previous section, reobtaining the differential equation deduced in ref. \cite{Hoenselaers} in a slightly different manner.

As explained in the introduction, it is not necessary to take the signal which guides the pirates  to be an electromagnetic wave (thus traveling at the speed of light), but that will be done here for the sake of definiteness. 
One could imagine that the pirates and the merchants are now in spacecrafts moving through outer space at speeds comparable to the speed of light. With that in mind, one could call the pursuers ``space pirates'' and their spacecraft trajectory would be a ``relativistic'' pursuit curve. As a matter of fact, we will discuss to what extent it makes sense to refer to those pursuit curves as ``relativistic" in section \ref{sec.rel}, but let us ignore those subtleties for now.

\begin{figure}[h!]
    \centering
    \includegraphics[width=9cm]{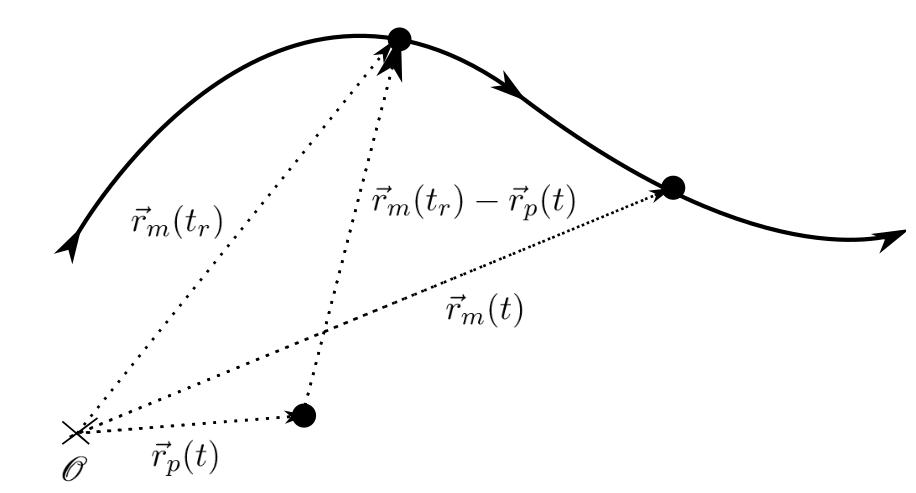}
    \caption{The concept of retarded time. It takes a finite amount of time ($t-t_r$) for the signal to leave the merchant spacecraft and reach the pirates. During that time interval, the merchants travel from position $\Vec{r}_m(t_r)$ to position $\Vec{r}_m(t)$. The pirates' velocity vector at time $t$ has the direction of $\Vec{r}_m(t_r) - \Vec{r}_p(t)$, where $\Vec{r}_p(t)$ denotes the pirates' position at that time.}
    \label{fig.retarded}
\end{figure}

At any given time, instead of pointing in the direction of the merchant spacecraft position at that time, the velocity vector of the pirate spacecraft points in the direction in which the pirates see the merchant spacecraft. Since it takes some time for the signal to reach the pirates after being emitted by the merchants (since its propagation speed is finite), those directions are in general not the same. This is illustrated in figure \ref{fig.retarded}.

As can be seen from  figure \ref{fig.retarded}, the signal detected by the pirates at time $t$ --- when their position is given by the vector $\Vec{r}_p(t)$ --- is not the one emitted by the merchant spacecraft at that time, but the one emitted some time before, when the merchants' position was given by $\Vec{r}_m(t_r)$. If the signal propagates at speed $c$, then the amount of time elapsed between emission and detection of the signal, $t-t_r$, satisfies the following equation:
\begin{equation}
     c (t - t_r ) = |\Vec{r}_m (t_r ) - \Vec{r}_p (t)| \,.
     \label{eq.retarded}
\end{equation}
Equation (\ref{eq.retarded}) can be regarded as an implicit definition of $t_r$, which is called ``retarded time.''

We henceforth particularize the discussion to the case of a merchant spacecraft traveling along a straight line, in order to compare it with the original problem analyzed in the previous section. As in that problem, both the merchants and the pirates travel at constant speeds, given respectively by $V_m$ and $V_p$, and we take the merchant spacecraft to move along the line $x=x_0$.

\begin{figure}[h!]
    \centering
    \includegraphics[width=6.5cm]{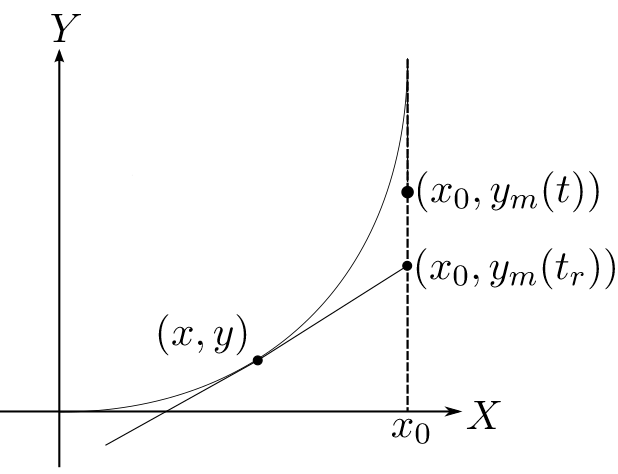}
    \caption{Pursuit curve for the case of a signal propagating at a finite speed. The pirates make for the point at which they see the merchant spacecraft, i.e. its position at retarded time. }
    \label{fig.rel}
\end{figure}

Moreover, we assume that the merchant spacecraft only becomes visible to the pirates once it crosses the $X$-axis of our coordinate system (one could imagine some big asteroid was blocking the view before that). We define $t=0$ to be the time at which the signal emitted by the merchants first reaches the pirate spacecraft. At that time, the pirate spacecraft is located at the origin of our coordinate system. Since the first signal had to travel a distance $x_0$ in order to reach the pirates, taking an amount of time $x_0/c$ to do so, the position of the merchant spacecraft at $t=0$ is given by $(x_0,V_m x_0/c)$. Therefore, at time $t\in \mathbb{R}$ the merchant spacecraft is located at the point $(x_0,y_m(t))$, with $y_m(t)=V_m x_0/c + V_mt$. Figure~\ref{fig.rel}  illustrates this situation for some $t > 0$, when the pirate spacecraft is located at an arbitrary point $(x,y)$.

Note from figure \ref{fig.rel} that the line which is tangent to the ``relativistic'' pursuit curve at  $(x,y)$ connects that point to $(x_0,y_m(t_r))$, i.e. the position of the merchant spacecraft at retarded time $t_r$, as opposed to its actual position $(x_0,y_m(t))$. This is the crucial difference between this problem and the one posed and solved by Bouguer.  We thus obtain an equation similar to (\ref{eq.class.slope}):
\begin{equation}
    \frac{dy}{dx} = \frac{y_{m}(t_r) - y}{x_{0} - x}= \frac{V_m x_0/c + V_mt_r - y}{x_0 -x}\,.
    \label{eq.rel.slope}
\end{equation}
Furthermore, equation (\ref{eq.class.arc}) is also valid here, for the same reasons given above it. 


The last piece of information we need comes from (\ref{eq.retarded}). In our case it becomes
\begin{equation}
    c(t-t_r) = \sqrt{(x_0-x)^2 + (y_m(t_r)-y)^2}\,.
\end{equation}
Now, from the first equality in (\ref{eq.rel.slope}) we have $y_m(t_r)-y = (x_0-x)dy/dx$, so
\begin{equation}
    c(t-t_r) = (x_0-x)\sqrt{1 + \left(\frac{dy}{dx} \right)^{2}}\,.
\label{eq.rel.ret}
\end{equation}

Finally, solving (\ref{eq.rel.slope}) for $t_r$ and (\ref{eq.class.arc}) for $t$, and then plugging the solutions into (\ref{eq.rel.ret}), we arrive at
\begin{equation}
    (x_0  -x)\sqrt{1 + p(x)^2} = \frac{c}{V_{p}}\int_{0}^{x}\sqrt{1+p(\xi)^{2}}d\xi - \frac{c}{V_{m}}(x_0 - x)p(x) - \frac{c}{V_m}y - x_0\,,
\end{equation}
where as before $p(x):=dy/dx$. Differentiating the above equation  with respect to $x$, we obtain once again a first-order differential equation for $p(x)$:
\begin{equation}
   \alpha\beta(x_0 - x)p(x)\frac{dp}{dx} = \alpha(1+\beta)(1+p(x)^2) - (x_0 - x)\sqrt{1+p(x)^{2}}\,\frac{dp}{dx}\,,
    \label{eq.rel.p}
\end{equation}
where we have defined $\beta:=V_p/c$ and we recall that $\alpha := V_m/V_p$. This equation is equivalent to equation (10) in ref. \cite{Hoenselaers}. Note that when $\beta \to 0$ it reduces to (\ref{eq.class.p}), as expected.

\section{Numerical results} \label{numerical}

Equation (\ref{eq.rel.p}) cannot be analytically solved for $p(x)$. To see why, note that one can integrate it, with the initial condition $p(0)=0$, to find\cite{typo}
\begin{equation}
    (1+p(x)^2)^{\alpha\beta/2}\left(p(x)+\sqrt{1+p(x)^2}\right) = \left(1 - \frac{x}{x_0}\right)^{-\alpha(1+\beta)}\,.
    \label{eq.rel.pint}
\end{equation}

\noindent This yields $x$ as a function of $p$, but it is not possible to invert that function.

Nonetheless, equation (\ref{eq.rel.pint}) is useful  in applying numerical methods to plot $y$ as a function of $x$ (see the Appendix for the precise Mathematica routine that we have used). We have done so for different values of the parameters $\alpha$ and $\beta$, and our results are shown in the figures below. Each graph includes:

\begin{itemize}
    \item the numerical solution corresponding to a given pair $(\alpha,\beta)$ (the thicker curve);
    \item the analytical solution to the original problem (equation (\ref{eq.class.sol})) for the same value of $\alpha$ used in the numerical solution (the thinner curve);
    \item a vertical dashed line corresponding to the trajectory of the merchant vessel/spacecraft;
    \item two horizontal dotted lines indicating the $y$-coordinates of the points at which the pirates reach the merchants in each version of the problem.
    
\end{itemize}
Note that the smaller the parameter $\beta$, the more similar the two curves are, as expected.

\begin{figure}[h!]
    \centering
    \includegraphics[width=7.5cm]{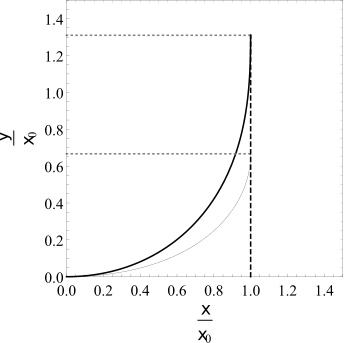}
    \caption{Pursuit curve for $\alpha=0.5$, $\beta=0.8$.}
    \label{fig.05_08}
\end{figure}

\begin{figure}[h!]
    \centering
    \includegraphics[width=7.5cm]{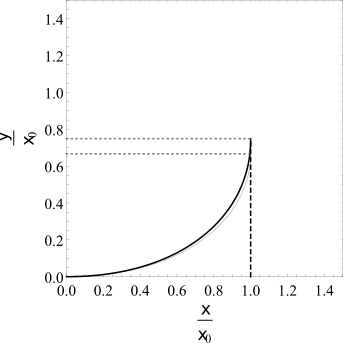}
    \caption{Pursuit curve for $\alpha=0.5$, $\beta=0.1$.}
    \label{fig.05_01}
\end{figure}

\begin{figure}[h!]
    \centering
    \includegraphics[width=7.5cm]{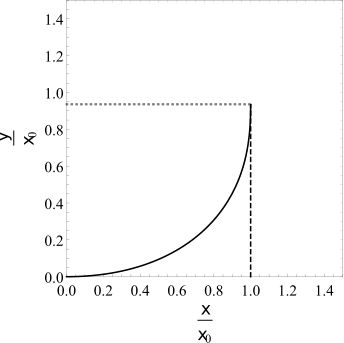}
    \caption{Pursuit curve for $\alpha=0.6$, $\beta=0.01$. Note the two curves are indistinguishable at this scale.}
    \label{fig.06_001}
\end{figure}

\begin{figure}[h!]
    \centering
    \includegraphics[width=7.5cm]{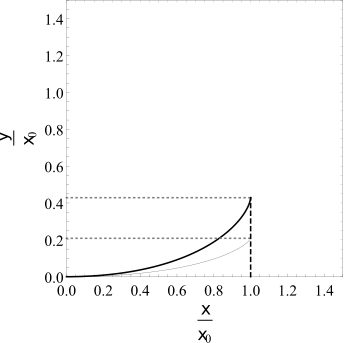}
    \caption{Pursuit curve for $\alpha=0.2$, $\beta=0.99$. }
    \label{fig.02_099}
\end{figure}

\section{Do the pirates ever reach the merchants?}

The finite speed of propagation of the signal emitted by the merchants  implies that the pirates' velocity vector does not point in the direction of the merchant spacecraft position at any given time, as illustrated in figure \ref{fig.rel}. This might suggest that the spacecrafts never actually meet. Is that really the case?

As mentioned below (\ref{eq.class.sol}), in the original problem a necessary and sufficient condition for the pirate ship to reach the merchant vessel is that the speed of the latter be lower than that of the former, i.e. $\alpha<1$. Otherwise the merchants escape.

Even though one cannot solve the ``relativistic'' problem exactly, it is possible to reach a conclusion on this matter as follows. First, note that, from the setup of the problem (or just by looking at the numerical plots in the previous section), it is safe to assume that the slope  $dy/dx$ becomes very large as the pirates approach the merchant spacecraft trajectory, i.e. for $x \approx x_0$. Therefore, we can take a certain value of $x$ in that region (as close to $x_0$ as we want) and approximate equation (\ref{eq.rel.pint}) considering $p(x)\gg 1$. Within this approximation, it is possible to solve the equation for $p(x)$, and the result is
\begin{equation}
   p(x) = \frac{dy}{dx} \approx \left({1 \over 2}\right)^{ 1 \over 1+ \alpha\beta}  \left(1 - \frac{x}{x_0}\right)^{-\frac{\alpha(1+\beta)}{1+\alpha\beta}}\,.
  \end{equation}
Integrating this equation from a certain value $x_\star$ to $x_0$, we find that the corresponding variation of the $y$ coordinate is proportional to the integral
\begin{equation}
\int_{x_\star}^{x_0}  \left(1 - \frac{x}{x_0}\right)^{-\frac{\alpha(1+\beta)}{1+\alpha\beta}} dx\,,
\end{equation}
which converges if and only if $\alpha < 1$. Hence, perhaps surprisingly, the necessary and sufficient condition for  the pursuer to catch the pursued is independent of $\beta$, and the same as in the original problem.

\section{Subtleties concerning the word ``relativistic''}\label{sec.rel}

Throughout this paper we have written the word ``relativistic'' this way, in between quotes. The reason is that  we actually use that word just to mean that information propagates at a finite speed, as was done by Hoenselaers.\cite{Hoenselaers}
In order to analyze the problem in a truly relativistic manner, we would have to take into account the fact that, in general, light rays are not straight lines in accelerated frames of reference,\cite{Scorgie} such as the pirate spacecraft. This implies we cannot really affirm that the pirates see the merchant spacecraft at its position at retarded time, i.e. equation (\ref{eq.rel.slope}) is not valid in general.

Of course, the discussion of the truly relativistic problem is outside the scope of this paper, since our goal is merely to introduce the concept of retarded time in a more friendly context. Nevertheless, it is worthwhile to investigate the conditions under which our results in the previous sections can be regarded as good approximations.

More precisely, we are interested in finding a constraint on the parameters $\alpha$ and $\beta$ such that the effects of acceleration can be considered small. Following ref. \cite{Scorgie}, we can expect  those effects to be small if the (proper) duration of the chase is much smaller than $c/A^{(0)}_p$, where $A^{(0)}_p$ is the magnitude of the pirates' proper acceleration. Since the speed is constant, the acceleration $A_p$ measured by the inertial observer of the previous sections is purely centripetal, so it can be estimated through the usual formula
\begin{equation}
    A_p = \frac{V_p^2}{R} \; \Longrightarrow\; A^{(0)}_p = (1-\beta^2)^{-1}A_p =  (1-\beta^2)^{-1}\frac{V_p^2}{R}\,,
    \label{eq.A}
\end{equation}
where $R$ is the radius of curvature of the trajectory at a given point and we have used the well-known relation between coordinate acceleration and proper acceleration.\cite{Schutz} Now, from calculus we know that
\begin{equation}
    R = \frac{\big[1 + (y^\prime(x))^2\big]^{3/2}}{|y^{\prime\prime}(x)|}\,.
\end{equation}
Plugging the solution of the classical problem (equation (\ref{eq.class.sol})) in the above expression, evaluated at $x=0$, so as to estimate the order of magnitude of $R$, we get $R\sim x_0/\alpha$, hence
\begin{equation}
    A^{(0)}_p  \sim (1-\beta^2)^{-1}\frac{\alpha V_p^2}{x_0}\,.
\end{equation}

On the other hand, it is easy to show that the duration of the chase, in the classical problem, is given by
\begin{equation}
    \Delta t = \frac{x_0/V_p}{1-\alpha^2}\,.
\end{equation}
Assuming that the duration of the chase in the relativistic case is of the same order as $\Delta t$, we can estimate the total proper time elapsed during the chase to be $\Delta\tau \sim \Delta t\sqrt{1-\beta^2}$. Therefore,
our results in the previous sections can be considered good approximations if
\begin{equation}
    \Delta \tau \ll \frac{c}{A^{(0)}_p} \; \Longleftrightarrow \; \frac{\alpha\beta}{(1-\alpha^2)\sqrt{1-\beta^2}} \ll 1\,.
\end{equation}

Finally, note that the considerations in this section are only relevant if we really take the signal which guides the
pirates to be an electromagnetic wave.  If we instead think of sound waves, with $c$ representing the speed of sound, then our results are equally good for any values of the parameters in the range $0\leq \alpha,\beta <1$.

\section{Final remarks and conclusions}

In this paper we have revisited the classical problem of pursuit, posed and solved by Bouguer in 1732, \cite{Pierre} taking into consideration that information cannot travel at infinite speed.
We have combined analytical and numerical methods in order to obtain the pursuit curve for different values of the parameters involved in the problem.

Crucial to the discussion was the concept of retarded time, which is usually introduced to students only in the context of electrodynamics, where it gets mixed with the complications inherent to the study of electromagnetic radiation. Interestingly, the Wikipedia has an entry on ``retarded time'' which is listed as one of the ``articles about electromagnetism.''
Here we have shown that the concept of retarded time can actually be introduced in the context of a problem which is essentially classical mechanical, hence much simpler. We believe that understanding retarded time in a context that is free from the complications of electromagnetism will reduce the difficulties students usually face when dealing with retarded potentials.

Furthermore, an important contribution of the present paper was to show that, as in the original problem, the necessary and sufficient condition for the pursuer to catch the pursued is that the speed of the latter be lower than that of the former, which might be a bit surprising, since it does not depend on the speed at which information travels.

There are a number of interesting directions in which one could try and extend the present work. Perhaps the most immediate one would be the investigation of this problem when the motion of the merchant spacecraft is not rectilinear. For instance, the case of a circular trajectory has been briefly discussed by Hoenselaers. \cite{Hoenselaers} But even without leaving the realm of rectilinear motion, there are a few questions worth thinking about. First, one could consider the problem we analyzed in this paper in the reference frame of the merchant spacecraft, which is inertial. This has been done for the original problem in ref. \cite{Silagadze}, where the authors found the pursuit curve in that frame. It would be very interesting to check if a similar analysis is possible in the ``relativistic" case.

Moreover, in the original problem one can show that there exists a constant of motion given by\cite{Silagadze2}
\begin{equation}
    C_0 := \frac{d}{dt}\big[\big(\vec{r}_m(t) - \vec{r}_p(t)\big)\cdot \big(\vec{v}_m(t) + \vec{v}_p(t)\big)\big] = V_m^2 - V_p^2\,,
    \label{C0}
\end{equation}
in the language of section \ref{s.retarded}, with $\vec{v}=d\vec{r}/dt$. Therefore, one can easily obtain the duration of the chase by integrating equation (\ref{C0}). Would it be possible to find an analogous conserved quantity 
 in the version of the problem involving retarded time? Since, in that case, $\vec{v}_p(t)$ is parallel to $\vec{r}_m(t_r) - \vec{r}_p(t)$, one could think of the following candidate:
 \begin{equation}
     C_\beta \stackrel{?}{=}  \frac{d}{dt}\big[\big(\vec{r}_m(t_r) - \vec{r}_p(t)\big)\cdot \big(\vec{v}_m(t_r) + \vec{v}_p(t)\big)\big].
 \end{equation}
 However, it is straightforward to verify that $C_\beta$ is not a constant of motion. This is related to the fact that, for the pirates, it appears as if the merchant spacecraft moved with variable speed, a fact which was also noticed by Hoenselaers.\cite{Hoenselaers}
 We currently do not know whether or not a generalized version of $C_0$ exists, so this fascinating problem remains open for future investigation.

\vskip 1 cm

\noindent
{\bf Acknowledgements}\\

We would like to thank Carlos Farina and Reinaldo de Melo e Souza for useful discussions and for valuable comments on the draft. Moreover, we thank Patr\'icia Abrantes and Daniela Szilard for their help with Mathematica, as well as J\'ulia Alves for reading the manuscript. We also  thank the Brazilian funding agency CNPq for partial financial support.

\vskip 1 cm

\appendix

\noindent
{\bf Appendix}

In order to produce the numerical plots displayed in section~\ref{numerical}, we have used the following Mathematica routine. We begin by defining the function
{\tt $$ X [p\text{$\_$}]\text{:=} \;1-\frac{1}{\left(\left(\sqrt{p^2+1}+p\right) \left(p^2+1\right)^{\frac{\alpha  \beta }{2}}\right)^{\frac{1}{\alpha  (\beta +1)}}}\,,$$}

\noindent which comes from solving  equation (\ref{eq.rel.pint}) for $x$ and taking $x_0=1$. Now we need to invert this function. Numerically, this means that we can make a table of points $(X, p)$ and then interpolate between those points. In Mathematica, this can be implemented as follows:
{\tt $$\text{graph = Table}[\{X[p],p\},\{p,0,100,0.1\}]; \qquad
\text{P = Interpolation[graph]} \,,$$}

\noindent so that the function {\tt P} is the numerical version of $p(x)$. But $p(x)$ is just $dy/dx$, so it suffices to numerically integrate the function {\tt P} to obtain $y$ as a function of $x$, i.e.
{\tt $$ \text{Y = Integrate[P[x],x]}\,. $$}

\noindent This is the function we plot to produce the figures in section \ref{numerical}.

A few comments are in order. Note that, in the function {\tt Table}, we have chosen the maximum value of $p$ to be 100 (which should correspond to a value of $x$ very close to $x_0$) and the increment $\Delta p$ as 0.1. Those choices yield reliable results when we set the value of the parameter $\alpha$ between 0.2 and 0.5. For values of $\alpha$ outside that range, it may be necessary to change the parameters in {\tt Table} in order to get a reliable result.

How can we check whether or not a given result is reliable? One way to do so is by computing the ratio between the total distance traveled by the pirates and the total distance traveled by the merchants, from $t=0$ till the capture of the latter. Defining $Y_{max}$ as the $y$-coordinate of the point where the capture takes place (numerically, {\tt $Y_{max} \text{ = } \text{Y} /. \;\text{x} \to$ 1.0}), that ratio is given by
$$
\Gamma := \frac{\int_0^1 \sqrt{1+p(x)^2}\,dx}{Y_{max}-\alpha\beta}\,.
$$
Now, since the speeds are constant, that ratio should give $1/\alpha$. Therefore, for a given function {\tt P}, we can compute $ \Gamma$ numerically and compare it to $1/\alpha$. We consider the results to be reliable if $| \alpha\Gamma - 1 |\lesssim 3\% $.



\vskip 1 cm

\noindent


\begin{thebibliography}{99}
\bibitem{typo} It is worth noting that (\ref{eq.rel.pint}) is equivalent to equation (11) in ref. \cite{Hoenselaers}, although there is a typo in the latter (the exponent on the right-hand side should be $-(v^{-1}+c^{-1})$).
\end{thebibliography}
\end{document}